\documentclass[preprint,showpacs,preprintnumbers,amsmath,amssymb,nofootinbib,showkeys,aps,11pt]{revtex4}
\usepackage{epsfig}
\usepackage{bm}									% Bold math
\usepackage{dcolumn}								% Align table columns on decimal point
\usepackage{graphicx}								% Include figure files
\usepackage{hyperref}
\usepackage{latexsym}
\usepackage{color}
\usepackage{hyperref}
\usepackage[T1]{fontenc}
\usepackage{pslatex}
\begin{document}

\title{Testing the  cosmic conservation of photon number with type Ia supernovae and ages of old objects}

\author{J. F. Jesus$^{1,2}$\footnote{E-mail: jfjesus@itapeva.unesp.br}}
\author{R. F. L. Holanda$^{3,4,5}$\footnote{E-mail: holanda@uepb.edu.br}}
\author{M. A. Dantas$^6$\footnote{E-mail: aldinezdantas@uern.br}}

\affiliation{$^1$Universidade Estadual Paulista (Unesp), C\^ampus Experimental de Itapeva, 18409-010, Itapeva - SP, Brasil}
\affiliation{$^2$Universidade Estadual Paulista (Unesp), Faculdade de Engenharia, Guaratinguet\'a,
% Departamento de F\'isica e Qu\'imica,
% Av. Dr. Ariberto Pereira da Cunha 333,
12516-410, Guaratinguet\'a - SP, Brasil.}
\affiliation{$^3$Departamento de F\'{\i}sica, Universidade Federal de Sergipe, 49100-000, São Cristovao - SE, Brasil.}
\affiliation{$^4$Departamento de F\'{\i}sica, Universidade Federal de Campina Grande, 58429-900, Campina Grande - PB, Brasil.}
\affiliation{$^5$Departamento de F\'{\i}sica, Universidade Federal do Rio Grande do Norte, 59300-000, Natal - RN, Brasil.}
\affiliation{$^6$Departamento de F\'{\i}sica, Universidade Estadual do Rio Grande do Norte, 59625-620, Mossor\'o - RN, Brasil}

\date{\today}

\begin{abstract}

{  In this paper, we obtain luminosity distances  by using ages of 32 old passive galaxies distributed  over the redshift interval $0.11 < z < 1.84$ and test the  cosmic conservation of photon number by comparing them with  580 distance moduli of type Ia supernovae (SNe Ia) from the so-called Union 2.1 compilation. Our analyses are based on the fact that the method of obtaining ages of galaxies relies on the detailed shape of galaxy spectra but not on galaxy luminosity. Possible departures from  cosmic conservation of photon number is parametrized by $\tau(z) = 2 \epsilon z$  and $\tau(z) =  \epsilon z/(1+z)$ (for $\epsilon =0$ the conservation of photon number is recovered). We find $\epsilon=0.016^{+0.078}_{-0.075}$ from the first parametrization and $\epsilon=-0.18^{+0.25}_{-0.24}$ from the second parametrization, both limits at 95\% c.l. In this way, no significant departure from cosmic conservation of photon number is verified. In addition, by considering the total age as inferred from Planck (2015) analysis, we find the incubation time $t_{inc}=1.66\pm0.29$ Gyr and $t_{inc}=1.23\pm0.27$ Gyr at 68\% c.l. for each parametrization, respectively.}

\end{abstract}

%\pacs{98.80.-k, 98.80.Es, 98.65.Cw}

\maketitle

\section{Introduction}
Since 1998, type Ia supernovae (SNe Ia) observations (Riess {\it et al.} 1998, Perlmutter {\it et al.} 1998, Suzuki {\it et al.} 2012, Bertoule et al. 2014) have been an important tool to access the current cosmic acceleration and test different cosmological models. In order to explain cosmic acceleration, if one wants to preserve Einstein's General Relativity Equations together with spacetime isotropy and homogeneity, one has to postulate some source of negative pressure. The sources of negative pressure that have been hypothesized include Einstein's cosmological constant (Padmanabhan 2003; Frieman, Turner \& Huterer 2008; Weinberg 2013), dark energy (Lima 2004; Caldwell \& Kamionkowski 2009; Li {\it et al.} 2011; Frieman, Turner \& Huterer 2008; Weinberg 2013) and quantum matter creation from gravitational field (Lima {\it et al.} 2008; Steigman {\it et al.} 2009; Lima {\it et al.} 2010; Graef {\it et al.} 2014; Jesus \& Andrade-Oliveira 2016).

However, this important evidence arising from SNe Ia observations has been questioned through the years and some alternative explanations to the observations have been given. Examples are possible evolutionary effects in SNe Ia events (Drell, Loredo \&  Wasserman 2000; Combes 2004), local Hubble bubble (Zehavi {\it et al.} 1998; Conley {\it et al.} 2007), modified gravity (Ishak, Upadhye \& Spergel 2006; Kunz \& Sapone 2007; Bertschinger \& Zukin 2008), unclustered sources of light attenuation (Aguirre 1999; Rowan-Robinson 2002; Goobar, Bergstrom \&  Mortsell 2002) and the existence of Axion-Like-Particles (ALPs), arising in a wide range of well-motivated high-energy physics scenarios, and that could lead to the dimming of SNe Ia brightness (Avgoustidis {\it et al.} 2009, 2010). Nowadays, the dark energy is supported by several independent observational data, such as baryon acoustic oscillations (BAO) and other galaxy clusters observations,  cosmic background radiation, observational Hubble constant data as well as age of the Universe (see Frieman, Turner \& Huterer 2008 and Weinberg 2013). {   Besides an accelerated stage, the high $z$ observational data also indicate a decelerated phase for $z>1$, fundamental for the structure formation process to take place (Riess et al. 2004).} 

With more than 700 Type Ia supernovae discovered (Betoule {\it et al.} 2014), the constraints on cosmological parameters inferred from SNe Ia are now limited by systematic errors rather than by statistical errors.  An important systematic error source is the mapping of the cosmic opacity. The   SNe Ia observations are affected by at least four different sources of opacity, namely, the Milky Way, the hosting galaxy, intervening galaxies, and the Intergalactic Medium.  The opacity can also occur by extragalactic magnetic fields that can turn photons into unobserved particles (e.g. light axions, chameleons, gravitons, Kaluza-Klein modes) (Avgoustidis {\it et al.} 2009 \& 2010). Recently, an interesting result was obtained  by Lima, Cunha \& Zanchin (2011). These authors discussed two different scenarios with cosmic absorption and concluded that only if the cosmic opacity is fully negligible, the description of an accelerating Universe powered by dark energy or some alternative gravity theory must be invoked (see also Li {\it et al.} 2013).

An interesting way to test the quality of the SNe Ia data has been performed in recent years by confronting them with data sets which are cosmic opacity independent. For instance, Holanda, Carvalho \& Alcaniz (2013) as well as Liao, Avgoustidis \& Zhengxiang (2015) used current measurements of the expansion rate $H(z)$ and SNe Ia data to impose cosmological model-independent constraints on cosmic opacity. These authors found that a completely transparent universe is in  agreement with the data considered (see also Avgoustidis {\it et al.} 2009, 2010 for analyses in a flat $\Lambda$CDM framework). Holanda \& Busti (2014) explored the possible existence of an opacity at higher redshifts ($z>2$) in $\Lambda$CDM context by using $H(z)$ data and luminosity distance of gamma ray bursts. The samples were compatible with a transparent universe at 1$\sigma$ level. Chen {\it et al.} (2012) by using  baryons acoustic oscillations (BAO) (Eisenstein et al. 2005) and SN Ia data  found that  an opaque universe is preferred in redshift regions $0.20-0.35$, $0.35-0.44$ and $0.60-0.73$, whereas a transparent universe is favoured in redshift regions $0.106-0.20$, $0.44-0.57$ and $0.57-0.60$. When these authors considered the entire redshift range, their result were still consistent with a transparent universe at $1\sigma$ confidence level (c.l.).  By testing the luminosity distance of SNe Ia with ADD from galaxy clusters, Li {\it et al.} (2013) have put constraints on cosmic opacity. However, the limits thus obtained rely on the assumptions used to describe the  galaxy clusters morphology. The results of  Holanda, Carvalho \& Alcaniz (2013) also rely on the SNe Ia light curve fitter (SALT2, MLCS2K2). In this way, it is still  interesting to investigate and compare different observational data and look for any systematics in them.

{  On the other hand, other interesting results  have arisen from this kind of analysis, with an excessive brightness being also detected in SNe Ia data. In this context, Nair, Jhingan \& Deepak (2012) compared distance measurements  obtained from SNe Ia and BAO and they found that the supernovae are brighter than expected from BAO measurements. Obtaining a similar result, Basset \&  Kunz (2004) found a $2\sigma$ violation caused by excess brightening of SNe Ia at $z>0.5$ when confronted them via cosmic distance duality relation with angular diameter distances (ADD) from compact objects, perhaps due to lensing magnification bias. However, such result also may come from some exotic sources of photon involving a coupling of photons to particles beyond the standard model of particle physics. An interesting effect occurs if there is mixing between photons and chameleons (a scalar particle which arises in certain models of scalar-tensor gravity) in presence of a strong magnetic field. In this context, observers on Earth  see a brightened image of the SNe Ia (see Burrage 2008 and Avgoustidis {\it et al.} 2010).}

{  In this paper, we show that  it is possible to test the  cosmic conservation of photon number with luminosity distances from SNe Ia and those inferred from 32 ages of old objects.  We do not assume any matter content in our analysis. Our analyses are based on the fact that the method of obtaining ages of galaxies relies on the detailed shape of galaxy spectra but not on galaxy luminosity. Moreover, it is only assumed that the Universe is homogeneous and isotropic, which leads to the Friedmann-Robertson-Walker geometry (see Eq. \ref{dc}).  As a simplifying hypothesis, we assume spatial flatness. In addition to the present analysis, in order to put constraints over $t_{inc}$, we use the {\it Planck} constraint on the cosmic total age (Ade  et al. 2016). As a result, no significant departure from cosmic conservation of photon number is verified.}

The paper is organized as follows. In Section II we describe our new method to verify the {  photon conservation}, while in Section III the observational quantities used in this work are discussed. The corresponding constraints on the opacity are investigated and discussed in Section IV. We summarize our main results in Section V.

\section{Methodology}
\subsection{Photon conservation and luminosity distances}
As well known, a {  violation of photon number from a luminosity source  leads to a modification of its inferred luminosity distance, increasing or decreasing it with respect to a transparent universe. Mathematically, if there is violation of photon number between observer and a light source, the flux received is modified by a factor $e^{-\tau(z)}$, where $\tau>0$ corresponds to a photon sink and $\tau<0$ corresponds to a photon source in the path to the observer % or increased by a factor $e^{\tau(z)}$
(Chen \& Kantowski, 2009a; 2009b). In this way, the inferred luminosity distance of the source, $D_{L,obs}$ is related to the true luminosity distance (in a transparent universe) by}
\begin{equation}
D^2_{L,obs}=D^2_{L,true}e^{\tau(z)}.
\end{equation}
Therefore, the observed distance modulus is given by
\begin{equation}
m_{obs}=m_{true} + 2.5[\log e]\tau(z).
\label{rela}
\end{equation}
In our analyses, measurements of $m_{obs}$ are taken from the SNe Ia Union 2.1 compilation (Suzuki {\it et al.} 2012). We compare $m_{obs}$ estimates from SNe Ia data to  $m_{true}$ inferred directly from the ages of old objects as will be discussed in the next subsection.

\begin{figure*}
\centering
\includegraphics[width=0.49\textwidth]{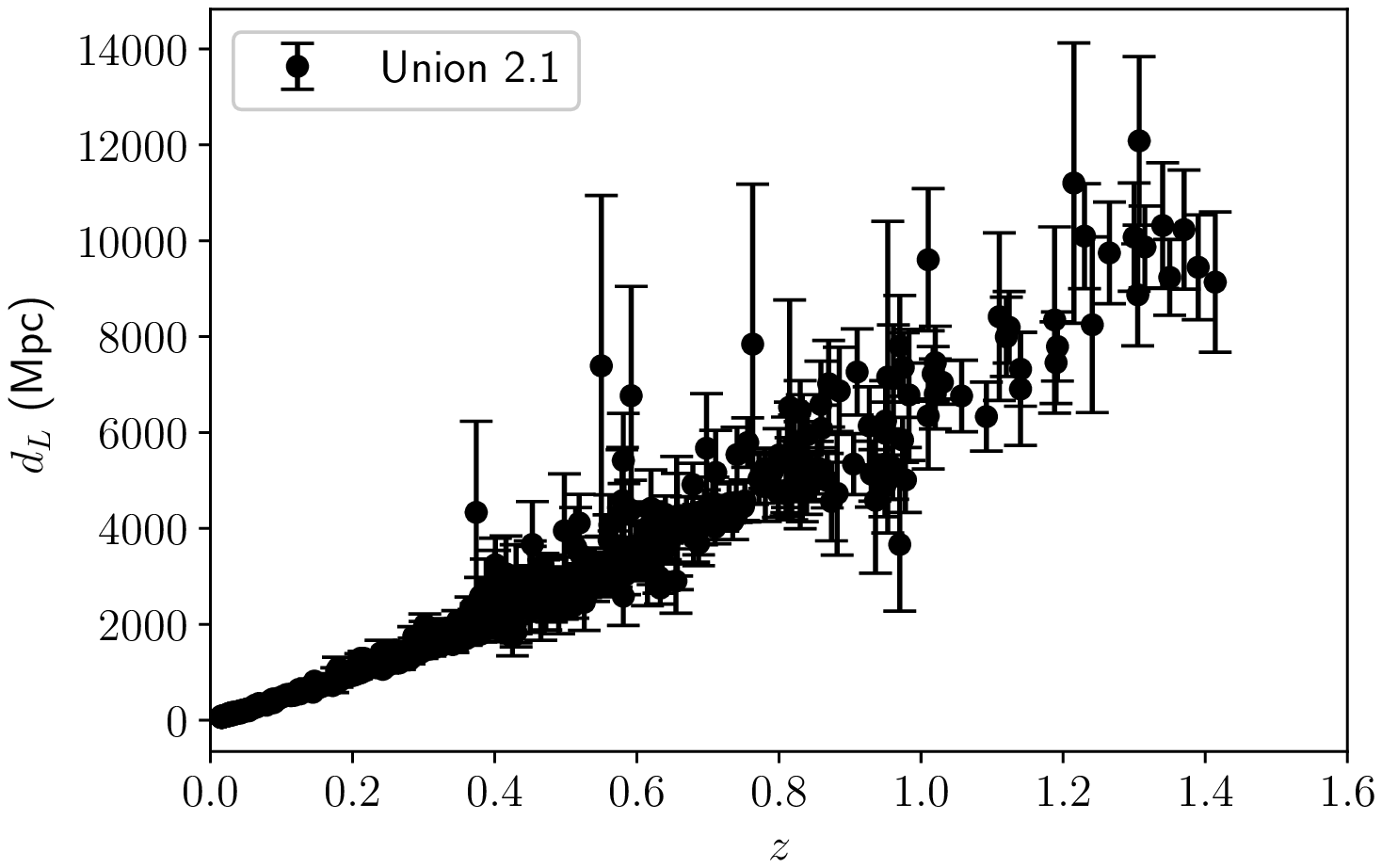}%{fig1.eps}
%\hspace{0.3cm}
\includegraphics[width=0.49\textwidth]{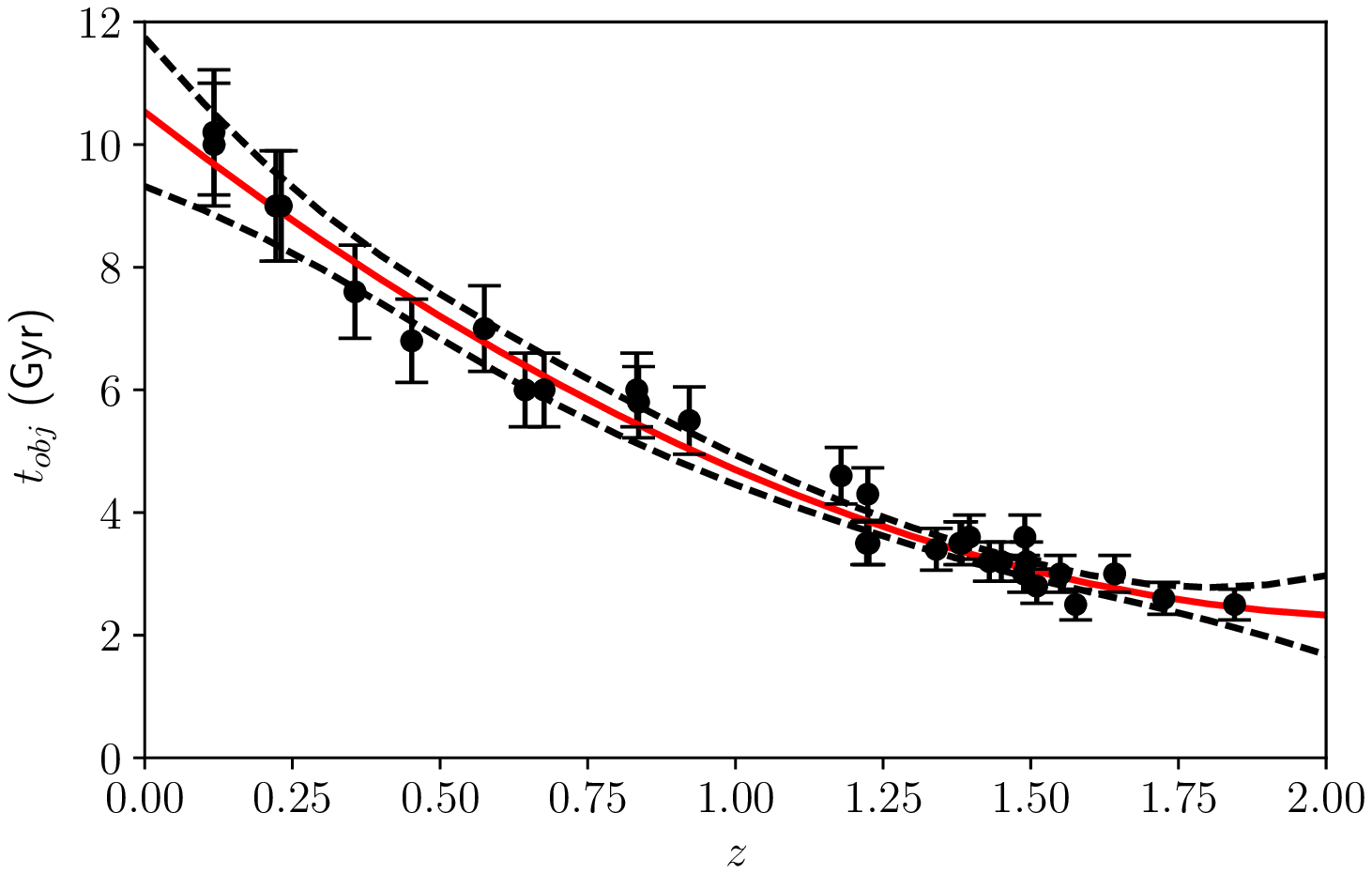}%{fig2.eps}
\caption{{  (a)} Luminosity distances of SNe Ia from Union 2.1 compilation. {  (b)} $t_{obj}(z)$ corresponding to the estimated ages of the galaxies.
%We plot the original data added with an incubation time $t_{inc} = 1.66$ Gyr.
The solid red and dashed black lines are the best fit and 2$\sigma$ error, respectively,  obtained from a fit by using a third degree polynomial.}
\label{figdata}
\end{figure*}

\subsection{Luminosity distance from old objects}

{  Let us assume the Friedmann-Robertson-Walker metric and that the method of obtaining ages of galaxies relies on the detailed shape of galaxy spectra but not on galaxy luminosity. Thus, the true luminosity distance can be obtained from ages of old objects as follows (Jimenez \& Loeb 2002)}
\begin{equation}
D_{L,true} = (1+z) c \int_z^0 \left[(1+z)\frac{dt}{dz}\right] dz,
\label{dc}
\end{equation}
where $t$ is the Universe age at redshift $z$. In this way, if one can access the $dt/dz$ quantity in SNe Ia redshift range without assumptions on cosmological model and also free of opacity, it is possible to obtain $m_{true}$ for each SNe Ia redshift and put constraints on $\tau(z)$ via Eq. (\ref{rela}).
% and put constrains on the cosmic opacity simply linking 

%\begin{equation}
%m_{true} = 5 log10 D_{L,true} + 25.
%\label{mtrue} 
%\end{equation}

In our work, $dt/dz$ is estimated from ages of 32 old passive galaxies distributed over the redshift interval $0.11 < z < 1.84$ (see next section for details). In Fig. 1b we plot the original estimated ages of galaxies (see next section for more details).
%As an estimate of total age of the universe ($z=0$), we use the most recent estimate from {\it Planck} results, $t_0=13.79 \pm 0.021$ Gy.
As we are only interested on the derivative $dt/dz$, instead of assuming an incubation time and using the total age from other observations, we choose to fit $t_{obj}(z)$. If we assume that the incubation time is constant, that is, independent of redshift, one may see that $t(z)$ only differs from $t_{obj}(z)$ by a constant. That is,
\begin{equation}
 t(z)=t_{obj}(z)+t_{inc}.
 \label{tztobj}
\end{equation}
We have tested some {  polynomial fits for $t_{obj}(z)$ and we have found that the minimal polynomial that yields a good fit, when combined with SNe Ia}, is a third degree polynomial fit, {  such as:
\begin{equation}
 t_{obj}(z)=a_0 + a_1z + a_2z^2 + a_3z^3.
 \label{tobjz}
\end{equation}

As we assume} $t_{inc}$ to be constant, we have $\frac{dt_{obj}}{dz}=\frac{dt}{dz}$, so we may say, that our model of Universe, that is, the function we assume that can describe the Universe given the data is:
\begin{equation}
 \frac{dt}{dz}=a_1+2a_2z+3a_3z^2
\end{equation}
%From this fit we obtain the following parameters (in Gyr): $A=13.8 \pm 0.01$, $B=-13.57 \pm 0.81$, $C=7.74 \pm 1.21$, $D=-1.79 \pm 0.45$, $\mathrm{cov}(B,C)=-0.25$, $\mathrm{cov}(B,D)=0.34$ and $\mathrm{cov}(C,D)=-0.54$ with reduced qui-square $\chi_{red}^2 = 0.38$.  Since we are interested in $\frac{dt}{dz} = B + 2Cz + 3Dz^2$, the covariances $\mathrm{cov}(A,B)$, $\mathrm{cov}(A,C)$ and $\mathrm{cov}(A,D)$ will not be used. 
From (\ref{tztobj}), we may also estimate $t_{inc}$, once we estimate $a_0$ and know $t_0$, as we have $t_{inc}=t_0-a_0$. Finally, from Eq. (\ref{dc}), the {  true luminosity distance} can be given by
\begin{equation}
D_L(z) = c(1+z)\left[-a_1\left(z+\frac{z^2}{2}\right)-a_2\left(z^2+\frac{2z^3}{3}\right)- a_3\left(z^3+3\frac{z^4}{4}\right) \right]. 
 \label{eqdc}
\end{equation}
However, one must note that the parameters $a_i$ derived from the age of old objects will be in Gyr, so in order to obtain $D_L$ in Mpc in Eq. (\ref{eqdc}) one must write $c$ in Mpc/Gyr as $c=306.6$ Mpc/Gyr.

%{  This conversion can be made as follows. By writing $c$ in km/s, we have
%\begin{equation}
% \frac{D_L}{\mathrm{Mpc}}=\frac{299792(1+z)}{\mathrm{km/s}}\frac{D_C}{\mathrm{Gyr}}\Rightarrow D_L(\mathrm{Mpc})=299792(1+z)\frac{\mathrm{Mpc}}{\mathrm{km}}\frac{\mathrm{s}}{\mathrm{Gyr}}D_C(\mathrm{Gyr})
%\end{equation}
%where $D_C(\mathrm{Gyr})$, the comoving distance in Gyr, is the expression in brackets in (\ref{eqdc}). So,
%\begin{equation}
% D_L(\mathrm{Mpc})=306.6(1+z)D_C(\mathrm{Gyr})
%\end{equation}
%}

\section{Data set}

In the following, we describe the data sets used in our analyses.

\begin{itemize}
\item {  For the photon number  dependent data}, we  use the Union 2.1 SNe Ia sample. This sample is an update of the original Union compilation (Amanullah {\it et al.} 2010) that comprises 580 data points including recent large samples from other surveys and uses SALT2 for SNe Ia lightcurve fitting (Guy {\it et al.} 2007). As the Union 2.1 consists of several subsamples, Suzuki {\it et al}. (2012) allowed a different absolute magnitude value for each subsample thereby making the impact of the cosmological model negligible.  {  This sample is plotted in  Fig. \ref{figdata}a, with the luminosity distances obtained via the known relation $D_L(z)=10^{(\mu(z)-25)/5}$, where $\mu$ is the 
distance modulus, and the corresponding $\sigma_{D_L}=\frac{d D_L}{d\mu}\sigma_\mu$.}

\item {For the photon number  independent data}, we use the age estimates of 32 old passive galaxies distributed over the redshift interval $0.11 < z < 1.84$, as recently analysed by Simon, Verde \& Jimenez (2005) {  (see Fig. \ref{figdata}b)}. The total sample is composed by three sub-samples: 10 field early-type galaxies from Treu {\it et al.} (1999, 2001, 2002), whose ages were obtained by using SPEED models of Jimenez {\it et al.} (2004); 20 red galaxies from the publicly released Gemini Deep Deep Survey (GDDS) whose  integrated  light  is  fully  dominated  by evolved  stars  (Abraham {\it et al.} 2004, McCarthy {\it et al.} 2004). Simon, Verde \& Jimenez (2005) re-analysed the GDDS old sample by using a different stellar population models and obtained ages within 0.1 Gyr of the GDDS collaboration estimates - and the two radio galaxies LBDS 53W091 and LBDS 53W069 (Dunlop {\it et al.}  1996; Spinrad {\it et al.}  1997; Nolan {\it et al.}  2001).

Recently, Wei et al. (2015) obtained $t_{inc} = 1.36$ Gyr and $t_{inc} = 1.62$ Gyr by considering this galaxy sample in a flat $\Lambda$CDM model with fixed and free $H_0$ (Hubble parameter) and $\Omega_m$ parameters, respectively. This factor accounts for our ignorance about the amount of time since the beginning of the structure formation in the Universe until the formation time of the object.  However, such a treatment assumes that  all of these galaxies need to have formed at the same time for their ages to trace out the Universe history. % In this way, we have added an average delay factor $t_{inc} = 1.50$ Gyr with a $30\%$ error such as $t_{inc} = 1.50 \pm 0.45$ covering the range found by Wei et al. (2015).
{We still assume $10\%$ uncertainty on measurement of age of each Galaxy (Dantas {\it et al}. 2009, 2011; Samushia {\it et al}. 2010)}. Again, it is also important to stress that the method for obtaining the age of galaxies relies on the detailed shape of galaxy spectra but not on galaxy luminosity, so it is independent of $\tau$ (Avgoustidis {\it et al.} 2010).

%\item  As total age ($z=0$) we have considered the most recent results from the {\it Planck} collaboration, such as $t_0= 13.79 \pm 0.021$ Gyr. This value was obtained from a jointly analysis by using cosmic microwave background radiation, SNe Ia, $H_0$ and the baryon acoustic oscillations (Ade {\it et al.} 2015).  
\end{itemize}

\section{Analyses and discussion}
{  We assume two possible departures from cosmic conservation of photon number}, as parametrized by two functions: 
\begin{itemize}
\item P1: $\tau(z)=2 \epsilon z$. This linear expression can be derived from the usual cosmic distance duality relation parametrization $D_L = D_A(1 + z)^{2+\epsilon}$ for small values of $\epsilon$ and $z \leq 1$, where $\epsilon$ quantifies departures from cosmic conservation of photon number (see Avgoustidis {\it et al.} 2009).
\item P2: $\tau(z)=\epsilon z/(1+z)$, which avoids the $\tau(z)$ divergence at high redshifts of the linear parametrization. 
\end{itemize}

{  As one may see,  $\epsilon > 0$ and $\epsilon < 0$ correspond, respectively, to the presence of a cosmic opacity or photon source between the observer and the light source.}

We estimate the best-fit to the set of parameters ${\mathbf{p}} \equiv (\epsilon,a_0,a_1,a_2,a_3)$ {  through a joint analysis involving the luminosity distances of SNe Ia and age of galaxies} by evaluating the likelihood distribution function, ${\cal{L}} \propto e^{-\chi^{2}/2}$, with 
\begin{eqnarray}
\chi^{2}(\mathbf{p}) & = & \chi^2_{tz}(a_0,a_1,a_2,a_3)+\chi^2_{SN}(\epsilon,a_1,a_2,a_3)
\label{chi2}
\end{eqnarray}
where
\begin{eqnarray}
\chi^{2}_{tz} & = & \sum_{i=1}^{32}\frac{(t_{obj,obs,i} - t_{obj}(z_i,a_0,a_1,a_2,a_3))^2}{\sigma^2_{t_{obj},obs,i}}
\label{chi2tz}
\end{eqnarray}
and
\begin{eqnarray}
\chi^{2}_{SN} & = & \sum_{i=1}^{580}\frac{(m_{obs,i} - m_{true}(z_i,a_1,a_2,a_3)-1.085736 \tau(z_i,\varepsilon))^2}{\sigma^2_{m,obs,i}},
\label{chi2sn}
\end{eqnarray}

{In Eq. (\ref{chi2tz}), $t_{obj}(z_i,a_0,a_1,a_2,a_3)$ is obtained from Eq.(\ref{tobjz}) and $t_{obj,obs,i}$ corresponds to ages of galaxies. In Eq. (\ref{chi2sn}), $m_{true}(z_i,a_1,a_2,a_3) = 5 \log_{10} D_{L,true} + 25$, with  $D_{L,true}$ obtained from Eq. (\ref{eqdc}). The quantities $m_{obs,i}$  and $\sigma^2_{m,obs,i}$ are the distance modulus and their uncertainties from SNe Ia, respectively.}%. where $m_{true} = 5 \log_{10} d_{L,true} + 25$ and $\sigma_{m(obs)}$ is the distance modulus and its uncertainty from SNe Ia.
%Following Suzuki {\it et al.} (2012) we added a 0.15 systematic error to SNe Ia data. Finally, the $\sigma^2_{m(true)}$ is given by:
%\begin{eqnarray}
%\sigma^2_{m(true)} & = & \left(\frac{\partial m_{true}}{\partial B}\right)^2\sigma^2_B +\left(\frac{\partial m_{true}}{\partial C}\right)^2\sigma^2_C  \nonumber\\ & + & 2\left(\frac{\partial m_{true}}{\partial B}\frac{\partial m_{true}}{\partial C}\right)\mathrm{cov}(B,C) \nonumber\\ & + &2\left(\frac{\partial m_{true}}{\partial B}\frac{\partial m_{true}}{\partial D}\right)\mathrm{cov}(B,D)\nonumber\\ & + & 2\left(\frac{\partial m_{true}}{\partial C}\frac{\partial m_{true}}{\partial D}\right)\mathrm{cov}(C,D).                                                               
%\end{eqnarray}

Because there is so many free parameters in both models, we choose to sample the likelihood through Monte Carlo Markov Chain (MCMC) analysis. A simple and powerful MCMC method is the so called Affine Invariant MCMC Ensemble Sampler by Goodman and Weare (2010), which was implemented in {\sffamily Python} language with the {\sffamily emcee} software by Foreman-Mackey {\it et al.} (2013). This MCMC method has the advantage over simple Metropolis-Hasting (MH) methods of depending on only one scale parameter of the proposal distribution and on the number of walkers. While MH methods in general depend on the parameter covariance matrix, that is, it depends on $n(n+1)/2$ tuning parameters, where $n$ is dimension of parameter space. The main idea of the Goodman-Weare affine-invariant sampler is the so called ``stretch move'', where the position (parameter vector in parameter space) of a walker (chain) is determined by the position of the other walkers. Foreman-Mackey {\it et al.} modified this method, in order to make it suitable for parallelization, by splitting the walkers in two groups, then the position of a walker in one group is determined by {\it only} the position of walkers of the other group\footnote{See Allison and Dunkley (2013) for a comparison among various MCMC sampling techniques.}.

\begin{figure*}
\centering
\includegraphics[width=\textwidth]{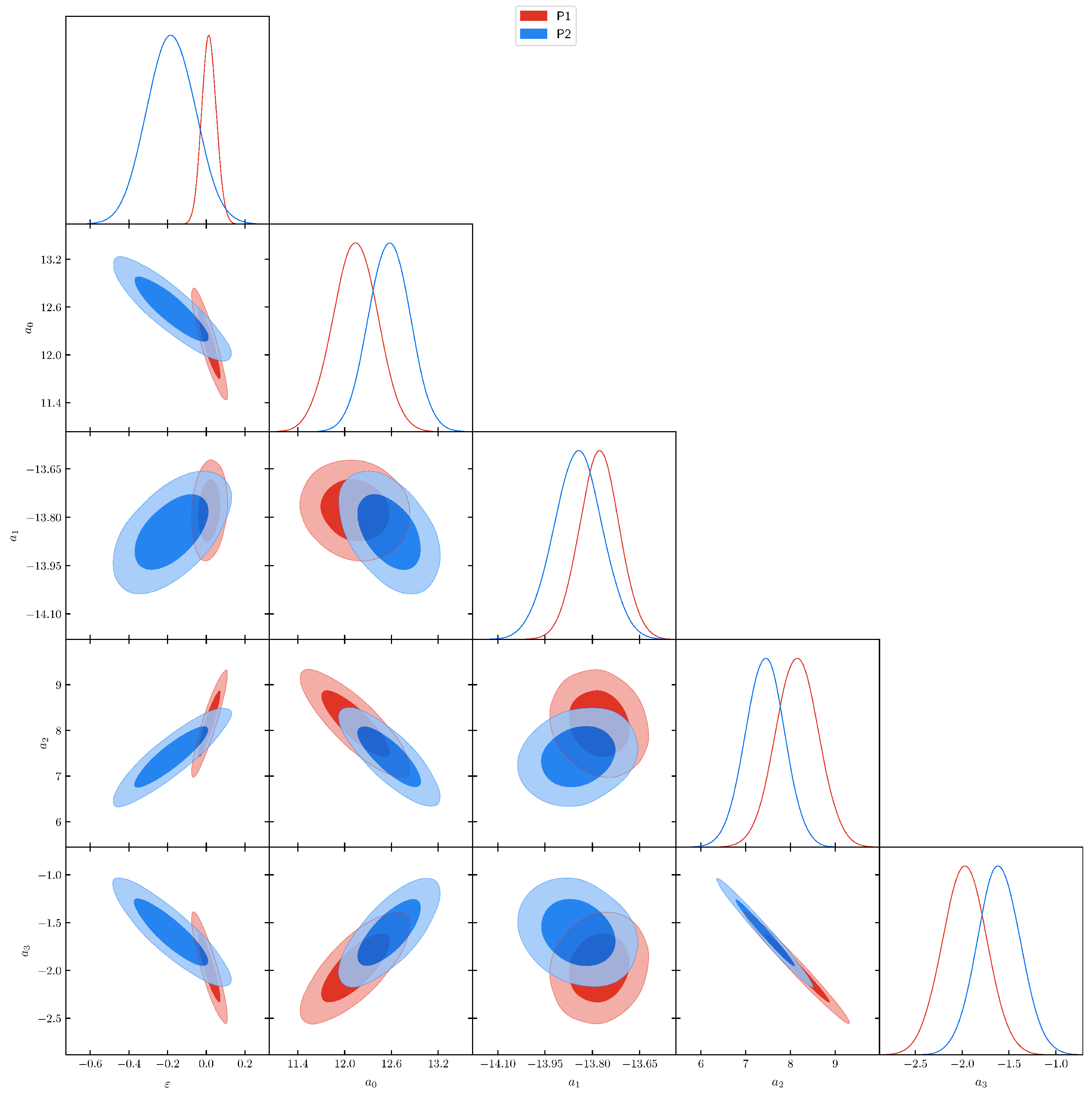}
\caption{The results of our statistical analysis, with constraints from SNe Ia Union 2.1 data and ages of old objects. {  Diagonal:} Likelihoods for the parameters on each indicated model, $P_1$ (red) and $P_2$ (blue). {  Below diagonal:} Contours for 68.3\% and 95.4\% confidence intervals for each indicated model, $P_1$ and $P_2$. The $a_i$ parameters have units of Gyr. Although we represent the $\epsilon$ parameter of each model in the same column, they have different meanings, because they correspond to different $\tau_i(z)$.}
\label{triangle}
\end{figure*}

We used the freely available software {\sffamily emcee} to sample from our likelihood in our 5-dimensional parameter space. We have used flat priors over the parameters.
%, as there are no physical bounds over them.
In order to plot all the constraints in the same figure, we have used the freely available software {\sffamily getdist}\footnote{{\sffamily getdist} is part of the great MCMC sampler and CMB power spectrum solver {\sffamily COSMOMC}, by Lewis and Bridle (2002).}, in its {\sffamily Python} version. The results of our statistical analyses from Eq. (\ref{chi2}) can be seen in Fig. \ref{triangle} and Table \ref{results}, {  where the errors correspond to 68.3\% and 95\% c.l.}

\begin{figure*}
\centering
\includegraphics[width=0.49\textwidth]{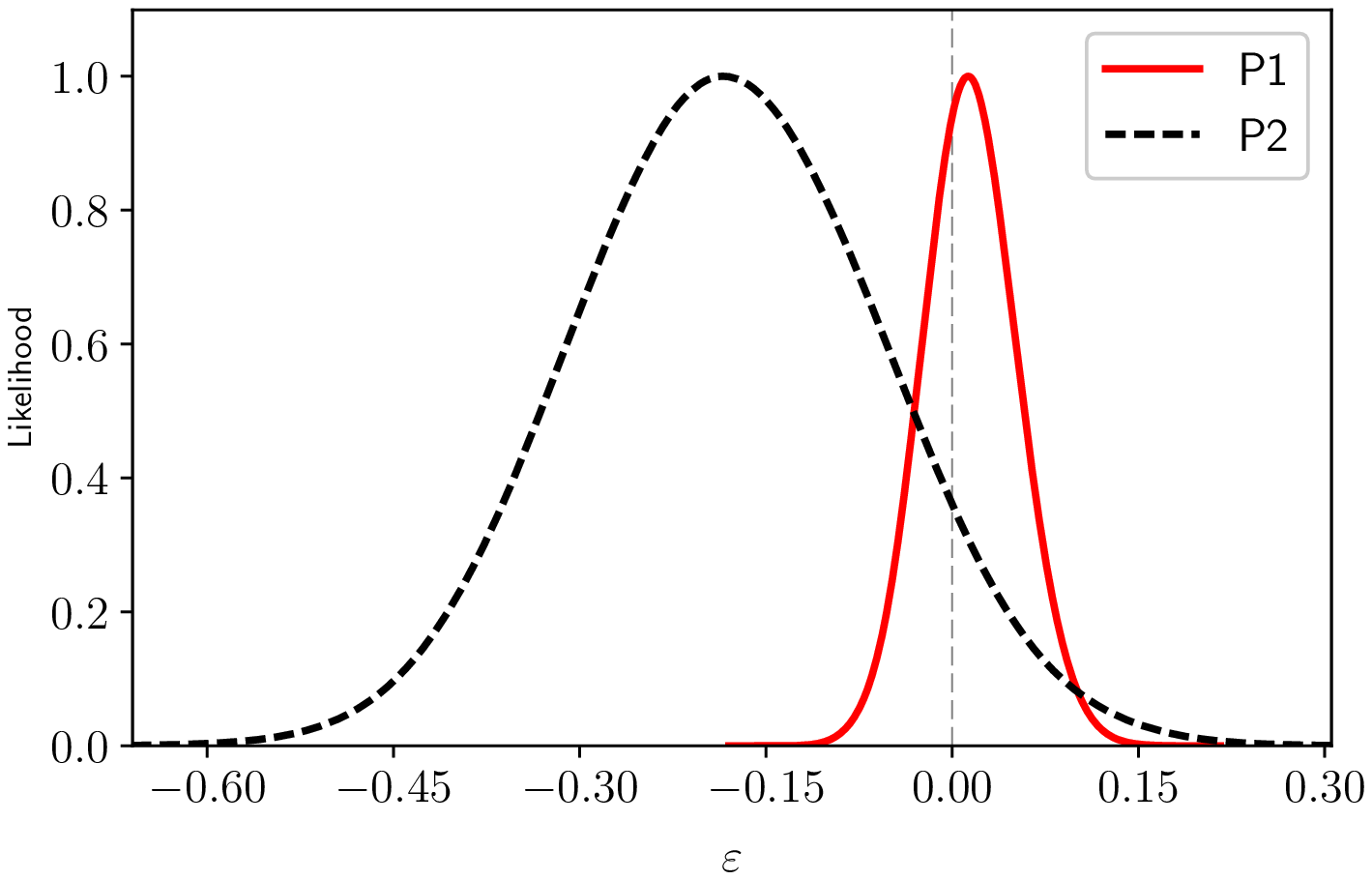}%{fig1.eps}
%\hspace{0.3cm}
\includegraphics[width=0.49\textwidth]{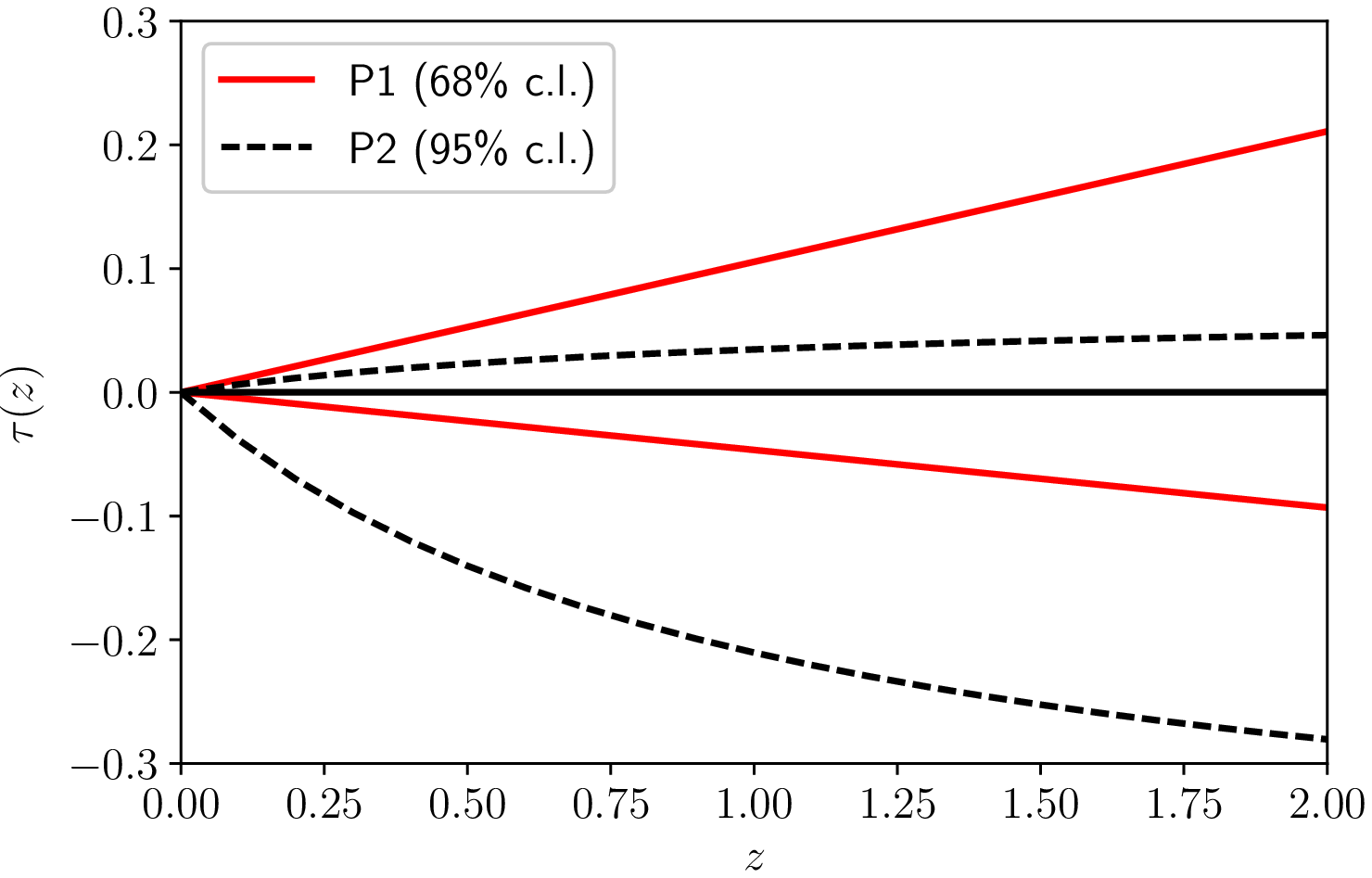}%{fig2.eps}
\caption{(a) Likelihood functions for $\epsilon$. (b) Confidence intervals on the plane $\tau - z$, 68\% c.l. for P1 and 95\% c.l. for P2. In both figures the solid and dashed lines correspond to results from P1 and P2.}
\label{liketau}
\end{figure*}

From Fig. \ref{triangle} and Table \ref{results}, we see that both functions (P1, P2) {  favour a cosmic conservation of photon number at least at 2$\sigma$ c.l. Figure \ref{liketau}a shows the marginalized likelihoods for $\epsilon$ in both models. An interesting result appears when the evolution of $\tau(z)$ is plotted. As one may see in Fig. \ref{liketau}b, the transparent universe, $\tau(z)=0$, is in full agreement with the data used in our analyses at 1$\sigma$ c.l. for model P1 and 2$\sigma$ c.l. for model P2.}

\begin{table}[t]
\begin{tabular} { c | c | c}
\hline
 Parameter &  P1 & P2\\
\hline
$\epsilon$      & $0.016^{+0.078}_{-0.075}       $& $-0.18^{+0.25}_{-0.24}            $\\

$a_0$ (Gyr)           & $12.14  \pm 0.58                   $& $12.57^{+0.54}_{-0.53}           $\\

$a_1$ (Gyr)           & $-13.778^{+0.13}_{-0.13}         $& $-13.845^{+0.15}_{-0.15}         $\\

$a_2$ (Gyr)          & $8.16^{+0.49+0.96}_{-0.48-0.96}             $& $7.43^{+0.86}_{-0.87}             $\\

$a_3$ (Gyr)           & $-1.98^{+0.48}_{-0.47}                    $& $-1.61  \pm 0.46           $\\
$\chi^2_{red}$           & $0.981            $& $0.978           $\\
\hline
\end{tabular}
\caption{Marginalized results for the free parameters of models P1 and P2. The central values shown correspond to mean values of the parameters and the errors correspond to 95\% c.l. The best fit values are much similar, as the distributions are quite symmetrical.}
\label{results}
\end{table}

As one may see on Table \ref{results}, the parameter errors are quite small, corresponding to 2.4\% (P1) and 2.1\% (P2) over $a_0$ (68.3\% c.l.), for example. This is due to the large number of SNe Ia and to the fact that the $t_{obj}(z)$ and SNe Ia are complementary, yielding nearly orthogonal constraints.

As we have claimed before, we may estimate incubation time in context of our models once we have estimated $a_0$ and by taking the total age from Planck (Ade {\it et al.}, 2016). The total age indicated by Ade {\it et al.} in their TT+lowP+lensing analysis, in the context of flat $\Lambda$CDM model, was $t_0=13.799\pm0.038$ Gyr. As $t_{inc}=t_0-a_0$, we have $t_{inc}=1.66\pm0.29$ Gyr for model P1 and $t_{inc}=1.23\pm0.27$ Gyr for model P2, both at 68\% c.l. These values are in agreement with the ones obtained by Wei {\it et al.} (2015), in the context of flat $\Lambda$CDM model, where they have found $t_{inc}=1.36$ Gyr for fixed and $t_{inc}=1.62$ Gyr for free $H_0$ and $\Omega_m$ parameters.

\subsection{Comparing results}

 At this point it is interesting to compare our results with previous ones that used the  linear parametrization  and different observations. 
For instance, Avgoustidis {\it et al.} (2009, 2010) via SNe Ia and $H(z)$ data obtained $\epsilon=-0.01^{+0.08}_{-0.09}$ and $\epsilon=-0.04^{+0.08}_{-0.07}$ 
in the flat $\Lambda$CDM framework. Holanda, Carvalho \& Alcaniz (2013) by using only SNe Ia + $H(z)$ observations obtained $\epsilon=0.017 \pm 0.055$. 
Holanda \& Busti (2014) by using  gamma ray bursts + $H(z)$ observations obtained $\epsilon=0.03 \pm 0.10$ and $\epsilon=0.028 \pm 0.10$ in flat $\Lambda$CDM and XCDM  
frameworks, respectively. More recently, Liao, Avgoustidis \& Li (2015) also used only SNe Ia + $H(z)$ data, but they have taken into account the covariance 
between the distances from $H(z)$ measurements obtained from integration on the $H(z)$ data, found $\epsilon = 0.07^{+0.11}_{-0.12}$, in full agreement with our results.  None of these analyses have been able to discard a transparent Universe ($\epsilon=0$).%On the other hand, these results also show that our assumption on $t_{inc}$ is completely realistic.

%found some strong evidence for the presence of a cosmic opacity. \\

\section{Conclusions}
\label{sec:conclusions}

In this paper we have proposed a new model cosmological independent method to probe the cosmic conservation of photon number. Although that the Universe acceleration {  for redshifts approximately lower than unity} is supported
by several other independent probes, investigating the cosmic opacity on the SNe Ia data is an important issue, in order to search for some source of 
unknown systematic error. If some extra dimming or brightness is still present, the SNe Ia observations will give us unreal values to main cosmological parameters 
and  the Universe will seem as accelerating at a different rate than it actually is.

To perform our analyses, we have considered the following cosmological data: 580 SNe Ia from Union 2.1 compilation and old objects, specifically, 32 old 
galaxies ($0.11 < z < 1.84$). {  Since the method to determine the ages relies on the detailed shapes of galaxy spectra but not on luminosities, they are independent of cosmic conservation of photon number. We have shown the possibility of obtaining  luminosity distances free of cosmic conservation of photon number assumption from the relation, in a flat FRW framework, between $D_{L}$ 
and $dt/dz$ quantity  from a best fit polynomial to $t(z)$ of old objects. Our ignorance about a possible departure from cosmic conservation of photon number was parametrized by 
$\tau(z)=2 \epsilon z$ (P1) and $\tau(z)=\epsilon z/(1+z)$ (P2) and we have found that $\epsilon$ is compatible with 0 at 1$\sigma$ c.l. for model P1 and at 2$\sigma$ c.l. for model P2 (see Fig. \ref{liketau}).
Thus, our results have reinforced the transparency of the universe and conservation of photons along with other analyses made in the literature, where were used SNe Ia, 
angular diameter distance and $H(z)$ data as well as have reinforced the present accelerated stage of the Universe.}

\begin{acknowledgments}
RFLH acknowledges financial support from INCT-A and CNPq (No. 478524/2013-7; 303734/2014-0). JFJ acknowledges financial support from FAPESP, Processes n$^\mathrm{o}$ 2013/26258-4 and 2017/05859-0, Funda\c{c}\~ao de Amparo \`a Pesquisa do Estado de S\~ao Paulo (FAPESP) and F. Andrade-Oliveira for helpful discussions.
\end{acknowledgments}

\end{document}